# Switching Mechanism in Single-Layer Molybdenum Disulfide Transistors: an Insight into Current Flow across Schottky Barriers

Han Liu,[†] Mengwei Si,[†] Yexin Deng,[†] Adam T. Neal,[†] Yuchen Du,[†] Sina Najmaei,[‡] Pulickel M. Ajayan,[‡] Jun Lou[‡] and Peide D. Ye[†,*]

[†] School of Electrical and Computer Engineering and Birck Nanotechnology Center, Purdue University, West Lafayette, IN 47907, USA

[‡] Department of Mechanical Engineering and Materials Science, Rice University, Houston, TX 77005, USA

Correspondence to: yep@purdue.edu

Abstract

In this article, we study the properties of metal contacts to single-layer molybdenum disulfide (MoS$_2$) crystals, revealing the nature of switching mechanism in MoS$_2$ transistors. On investigating transistor behavior as contact length changes, we find that the contact resistivity for metal/MoS$_2$ junctions is defined by contact area instead of contact width. The minimum gate dependent transfer length is ~0.63 μm in the on-state for metal (Ti) contacted single-layer MoS$_2$. These results reveal that MoS$_2$ transistors are Schottky barrier transistors, where the on/off states are switched by the tuning the Schottky barriers at contacts. The effective barrier heights for source and drain barriers are primarily controlled by gate and drain biases, respectively. We discuss the drain induced barrier narrowing effect for short channel devices, which may reduce the influence of large contact resistance for MoS$_2$ Schottky barrier transistors at the channel length scaling limit.

The rise of semiconducting 2D crystals has given much opportunity for future electronic and photonic devices.[1,2] High performance $MoS_2$ transistors, based on single-layer or multi-layer crystals, have been demonstrated with the following properties: reasonable electron mobility from several dozens to hundreds, high drive current, low sub-threshold swing, and superior immunity to short channel effects.[3-7] $MoS_2$ transistors offer several advantages over bulk semiconductor transistors. First, the atomically flat nature of $MoS_2$ leads to intrinsically low surface scatting, allowing the channel thickness to be scaled to the sub-nanometer regime. In contrast, the rough surface of ultrathin body (UTB) silicon would lead to severe surface scattering for carriers at this channel thickness. Second, besides its extremely thin body, the dielectric constant of $MoS_2$ is relatively low (~3.3), making it more robust against short channel effects than silicon.[7] To illustrate, a single-layer $MoS_2$ transistor (with body thickness of ~0.65 nm) with 1 nm equivalent oxide thickness (EOT) would yield a characteristic screening length of 0.7 nm, a surprisingly low number compared to transistors with conventional bulk semiconductors.[6,7] Third, large density of states, correlated to the high effective mass, would lead to high performance of the transistor at the scaling limit.[8]

However, to realize all the merits of $MoS_2$ transistors, there are still several technical issues and challenges ahead. In recent years of extensive studies on $MoS_2$, researchers have encountered two bottlenecks for further development of $MoS_2$ transistors: difficulty with dielectric integration and large contact resistance. The first issue, regarding dielectric integration, is key to achieving low EOT on top of $MoS_2$.[9] In the absence of dangling bonds on the crystal surface, dielectric growth relies on physical adsorption of atomic layer deposition (ALD) precursors. This physical adsorption process interferes with the self-limiting nature of the ALD process, making it challenging to form a defect-free low EOT dielectric on top. The second issue, large contact resistance, originates from the existence of the Schottky barrier at the metal/MoS$_2$ interface. Due to Fermi-level pinning, a Schottky barrier is evident at all metal/semiconductor interfaces.[10] A common approach to deal with this issue is to heavily dope the semiconductor, so that electrons can easily tunnel from the metal to the semiconductor. However, for MoS$_2$ transistors, no reliable doping technology, such as ion-implantation for bulk semiconductors, has been developed with fine control of doping concentration and doping profile.[11] Therefore, without the controlled heavy doping, the effect of Schottky barrier at the metal/MoS$_2$ must be considered. A previous study has revealed that the Fermi-level is pinned near conduction band edge, thus making MoS$_2$ transistors mostly n-type. Despite the barrier height for electrons being relatively small, ranging from 30 to 230 meV depending on back gate bias and the work function of contact metal,[6,12] the barrier has a profound impact on the device performance. The existence of the Schottky barrier causes the MoS$_2$ transistor to operate in a completely different manner from conventional Si MOSFETs. In this article, we take a deep look at the role of the Schottky barriers in MoS$_2$ transistors from the device perspective by studying the contact properties in single-layer MoS$_2$ transistors. Instead of calculating the Schottky barrier heights,[6] we determine the contact resistivity and transfer length for the junctions and see how they change with gate voltage. Our results show that the transfer length is inversely proportional to gate voltage, and this relationship reveals the nature of current flows across the junction and the switching mechanism in single-layer MoS$_2$ transistors. The device operation is dominantly controlled by tuning the effective heights of both Schottky barriers at source and drain contacts, instead of the potential barrier in the channel.

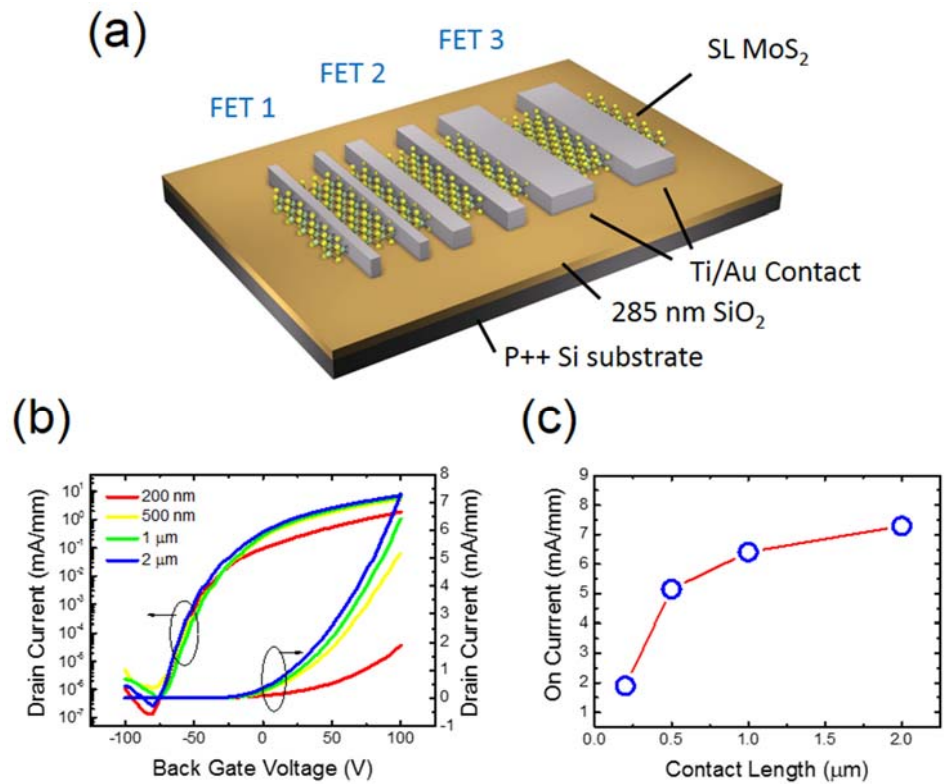

Figure 1: (a) Schematic view of device configurations. P++ silicon wafer capped with 285 nm SiO$_2$ was used as the global gate and gate dielectric, respectively. Single-layer MoS$_2$ films were grown *via* CVD methods and were etched to rectangular shapes with a uniform width of 2 μm. Ti/Au metal contact pairs with various contact length ranging from 0.2 to 2 μm were used as S/D contact metal. (b) Transfer curves for all devices with various contact length at 2 V drain bias. (c) Contact length dependent on-current at 100 V back gate bias.

In our experiment, single-layer MoS$_2$ crystals were grown using chemical vapor deposition (CVD) techniques.[13] The details of the material synthesis can be found in our previous publications.[14] The $MoS_2$ was grown on a heavily doped silicon wafer capped with 285 nm $SiO_2$, and most of the as-grown $MoS_2$ crystals appeared in triangular shapes. Prior to device fabrication, these triangular flakes were patterned by e-beam lithography and dry-etched by $BCl_3$/Ar plasma into rectangularly shaped channels all having a channel width of 2 μm. Contact bars with lengths of 0.2, 0.5, 1 and 2 μm were defined with e-beam lithography. Each pair of contact bars has a fixed spacing of 1.1 μm, thus we can use the long channel approximation in the following discussions of device performance. Ti/Au was used as the contact metal here as Ti is a low work function metal and has been used to create high performance $MoS_2$ transistors based on single-layer crystals.[14] No annealing was performed after the metallization process. Due to the 2D nature of single-layer $MoS_2$ crystals, we assume that the metal/$MoS_2$ interface is unperturbed by chemical reactions. The final device structure is illustrated in Figure 1(a). We use the global back gate to modulate these devices instead of the top gate because the global back gate can better modulate both carrier density in the channel and the Schottky barrier across the contact. This provides us a more direct view of how contact resistance and channel resistance change individually under the same gate voltage, in order to better reveal how the $MoS_2$ transistor operates at different bias conditions.

**RESULTS AND DISCUSSIONS**

Each pair of contact bars with same contact length was taken as the source/drain pair for an individual back-gated transistor. The transfer curves of devices measured at 2 V drain bias with various contact length are compared in Figure 1(b). All devices show clear switching behavior with a current on/off ratio over ~$10^6$. Similar threshold voltage ($V_T$) and sub-threshold swing (SS) are observed as well. Meanwhile, all devices show similar off-current level around $10^{-6}$~$10^{-7}$ mA/mm, regardless of the contact length. However, if we take a look at the on-current, there is an obvious change in the dependence of on-current with contact length. Contact length dependent on-current is shown in Figure 1(c). At 100 V back gate bias, the 0.2 μm contact length device has the on-current of ~2 mA/mm. If the contact length is expanded to 0.5 μm, the on-current increases up to ~5.1 mA/mm, almost proportional to the contact length. However, with further expansion of contact length, the current does not increase proportionally and gets saturated at ~7.3 mA/mm. This phenomenon is rarely seen in conventional Si MOSFETs at micron dimensions, where a low resistive contact impacts the on-current in only a minor way. We may roughly explain this phenomenon as a large contact resistance in MoS$_2$ transistors which accounts for a larger portion of the total device resistance. The large contact resistance was also observed in our previous study, where the drain current saturated at shorter channel lengths with channel length scaling.[6] However, a further examination of the contact resistance allows us to understand switching mechanism for MoS$_2$ transistors, as discussed in the later parts of this article.

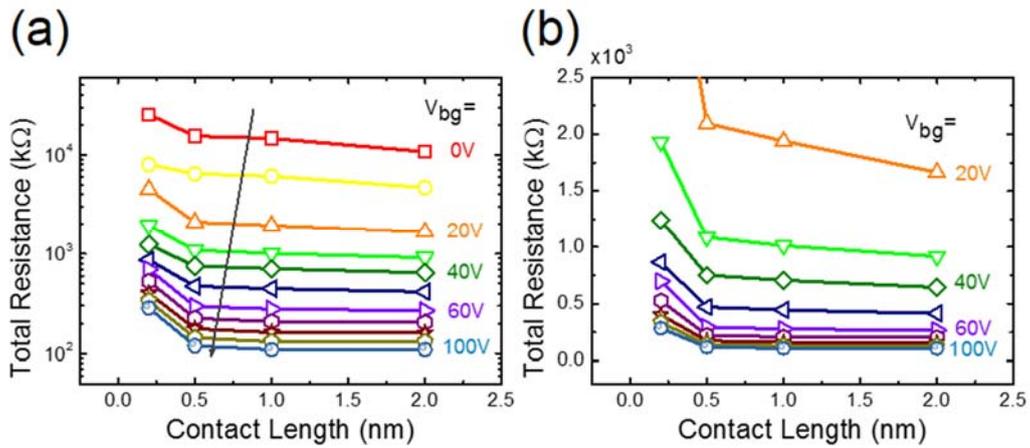

Figure 2: (a) Measured total resistance of the devices under various back gate bias with different contact length. A low drain bias of 50 mV was used for all measurements. (b) Linear scale of same data set of back gate bias larger than 20 V.

We extract the total resistances for different contact lengths at various back gate biases, as plotted in Figure 2, where a low drain bias of 50 mV was applied for all measurements. The back gate was biased from 0 V to 100 V. Clearly, by changing the gate voltage, the total resistance changes by two-orders of magnitude, independent of the contact length. Meanwhile, the impact of contact length on total resistance can be seen at each gate voltage. We notice that this impact is quite different at various gate voltages. For example, at zero gate bias, the total resistance for 0.2 μm contact is 25 MΩ and this resistance is reduced to 15, 14 and 10 MΩ when the contact length is increase to 0.5, 1 and 2 μm. This drop can be clearly seen in both log scale and linear scales shown in Figure 2. However, with further increase of the gate bias, the reduction of total resistance is not as strong as it was for low gate biases. For 100 V gate bias, the total resistances are 0.29, 0.12, 0.11 and 0.11 MΩ for 0.2, 0.5, 1 and 2 μm contact length, respectively. The total resistance seems to approach a minimum value at 0.5 μm, and the difference cannot be identified even with the log scale of Figure 2(a). Considering the two gate voltages discussed above, one sees that the dependence of the total device resistance on contact length changes as a function of gate voltage or, equivalently, the carrier density in the semiconductor at the metal/semiconductor junction. When the carrier density is low in the semiconductor, the total resistance depends more strongly on the contact length, but when carrier density is increased, the dependence weakens. This contact-length dependent transport behavior grants insight into the nature of current flow across the metal/atomically thin semiconductor junction and the operational mechanism of single-layer MoS$_2$ transistors at different bias conditions, both of which will be discussed in more detail below.

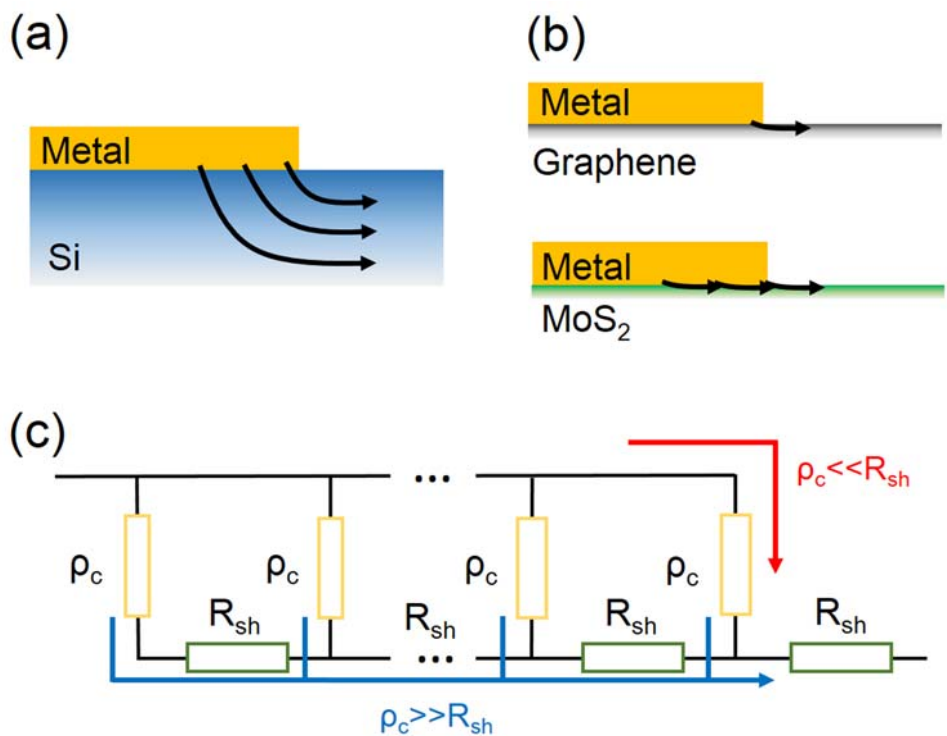

Figure 3: (a) Schematic view of current flow across the metal/bulk Si, where current flows deep inside the semiconductor. (b) Schematic view of current flow across metal/2D crystal. For metal/graphene junction, current flow crowds at contact edge.

However, for metal/MoS$_2$ junction, the transfer length is larger. (c) The resistor network model at the metal/semiconductor junction.

We first discuss how current flows across the metal/single-layer MoS$_2$ junction. Starting with the conventional metal/bulk silicon junction, as shown in Figure 3(a), the carriers injected into silicon from the metal contacts flow not only at the semiconductor surface but also deep into the bulk.[15] The depth of the current flow in this case is primarily determined by the junction depth of the implanted region of the semiconductor. Now considering two dimensional materials, since the penetration depth of current flow is limited by the body thickness when the bulk crystal is reduced to atomic thickness, an intuitive description of current flow across the junction would be that the current only flows across the junction at the contact edge, as shown in the top of Figure 3(b). This assumption was proved to be true in graphene.[16] It has been shown that the contact resistivity ($\rho_c$) of the metal/graphene junction is defined by $\rho_c = R_c \times W$ instead of $\rho_c = R_c \times A$, where $R_c$ is the contact resistance, $W$ is the contact width, and $A$ is the contact area. Though there are still some controversies on the graphene contact properties, experimental studies show similar results and large currents have been obtained with narrow metal contacts in graphene transistors. However, our results show that current flow strictly at the edge of the metal contact, as in graphene, may not hold true for other semiconducting 2D crystals. In the case of MoS$_2$, the carriers usually need a larger contact length to realize adequate carrier injection, depending on the gate bias, as shown schematically in the bottom of Figure 3(b). A resistor network is usually applied to model the metal/semiconductor junction, as shown in Figure 3(c). When current flows across the junction, it encounters two resistances. One is the impedance from the Schottky barrier, where it is simplified as a resistor $\rho_c$, and a sheet resistor $R_{sh}$.[16] The current would choose the least resistive path from the metal to the semiconductor. The potential distribution under the contact is determined by both resistors and can be written as:[17]

$$V(x) = \frac{I\sqrt{R_{sh}\rho_c}}{W} \frac{\cosh[(L-x)/L_T]}{\sinh(L/L_T)}$$

where $x$ is the lateral distance from the contact edge, $L$ and $W$ are the contact length and width, $I$ is the current flowing into the contact. The voltage is highest near the contact edge and drops nearly exponentially with distance. Usually, the "$1/e$" distance of the voltage drop is defined as the transfer length $L_T$ and can be expressed as $L_T = \sqrt{\rho_c/R_{sh}}$. In conventional Si MOSFETs, both $\rho_c$ and $R_{sh}$ are almost fixed numbers for implanted regions, while, in MoS$_2$ transistors, they are modulated by gate voltage. In the resistor network in Figure 3(c), if $\rho_c$ is much larger than $R_{sh}$, the least resistive path would be all routes in the network, marked as blue lines in Figure 3(c). This situation would correspond to infinite transfer length. In contrast, if $R_{sh}$ is much larger than $\rho_c$, then all current flows through the into the channel region at the junction edge, as noted in the red direction. In this case, the contact length would be very small. Metal contacts on graphene are best described by the second case, while MoS$_2$ lies in between. Both $R_{sh}$ and $\rho_c$ are modulated by the back gate bias and lead to the behavior of $L_T$ in MoS$_2$ which is discussed below.

Before discussing the mechanisms of single-layer MoS$_2$ transistor operation at different bias conditions, we need to know how $L_T$ changes with gate voltage. As we have shown in Figure 2(a) and (b), the $L_T$ is not a constant number in MoS$_2$ transistors, as both $R_{sh}$ and $\rho_c$ at the contact regions are dependent on the back gate bias. Actually, the drain bias has an impact on them as well, but we first look at how gate voltage changes the $L_T$. In order to determine $L_T$, we must calculate both $R_{sh}$ and $\rho_c$. To begin, we determine the $R_{sh}$, plotted in Figure 4(a), using the transmission line method (TLM) structure fabricated on the same sample. At 0 V back gate bias, the sheet resistance is ~3MΩ/□. With increasing gate bias, the sheet resistance reduces to ~106 KΩ/□. This can be easily understood as the Fermi level in MoS$_2$ is raised up by back gate biasing, inducing higher carrier density and hence reducing the sheet resistance.

With the sheet resistance determined, we then calculate and discuss R$_c$ as an intermediate step in determining $\rho_c$ and L$_T$. R$_c$ is calculated by subtracting the sheet resistance times the geometry factor from the total device resistance. The $R_c$ for the different contact lengths are plotted in Figure 4(b). The decreasing trend in $R_c$ can be attributed to the increasing carrier density in MoS$_2$ under the metal contacts. The higher carrier density induced by the electric field leads to a narrower Schottky barrier, facilitating thermal-assisted carrier tunneling to the semiconductor. In this case, the semiconductor can be viewed as being "electrostatically doped" by gate biasing. As expected, shorter contact length yields a higher $R_c$. As Figure 4(b) shows, the extracted R$_c$ depend more strongly on contact length for low gate voltage bias. As the gate voltage is increased, with the exception of the 0.2 μm contact length, the contact resistance of the other contact lengths become more and more similar. This behavior indicates that, at lower gate biasing, the contact metal needs a larger contact length to realize a full carrier injection. However, at high gate biasing, a smaller contact area is adequate.

The contact resistivity is calculated by $\rho_c=R_c\times A$ as mentioned above. At this point, it is important to choose the right contact dimension to estimate the contact resistivity. The 2 μm contact length shows 2 times larger contact resistivity than 0.2 μm contact length, as shown in Figure 4(c). This difference is primarily from larger potential drop along the contact length for the 2 μm contact as compared to 0.2 μm contact. This change in potential causes the contact resistivity calculated from larger contact length to be overestimated. The more precise transfer length $L_T=\sqrt{\rho_c/R_{sh}}$ is determined by using $\rho_c$ calculated from smaller contact dimensions as shown in Figure 4(d). At 0 V gate bias, the transfer length for Ti/Au contacts on single-layer MoS$_2$ is 1.26 μm, which drops to around 0.63 μm at the high gate biases. This suggests that, for single-layer MoS$_2$ transistors, the contact length should be at least ~1 μm (1.5L$_T$) to guarantee the least contact resistance when device is in the on-state.

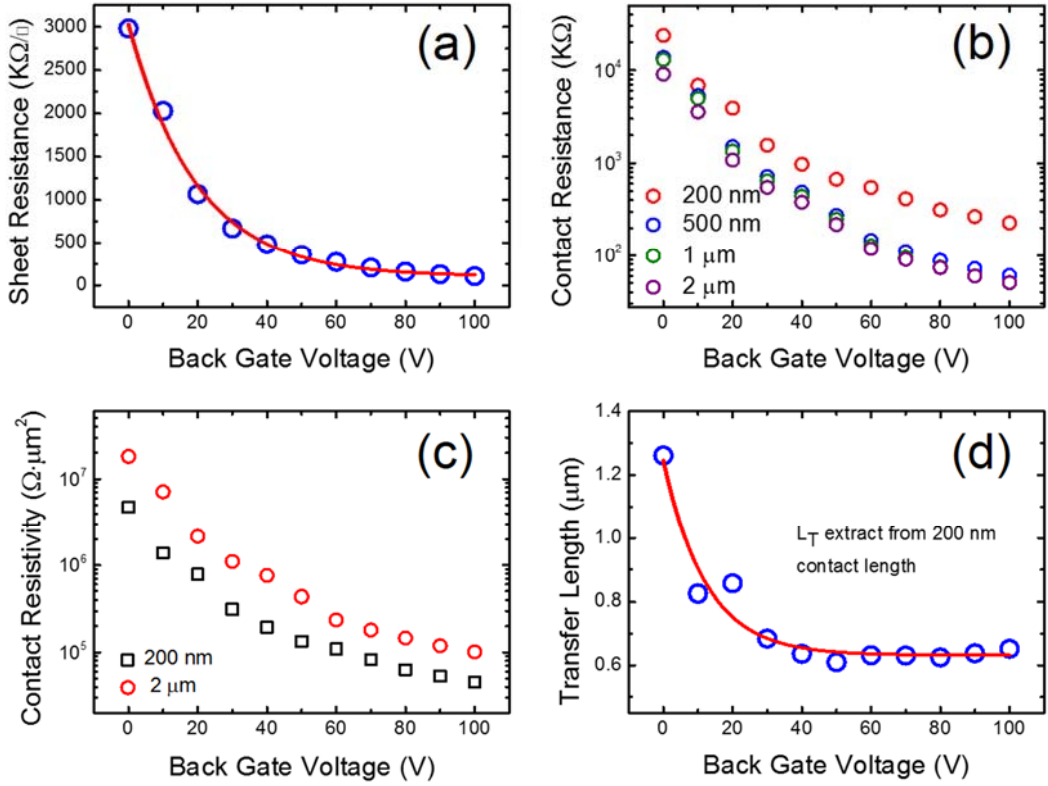

Figure 4: (a) Gate-voltage dependent sheet resistance ($R_{sh}$) estimated from TLM structure. (b) Contact resistance for different contact length at various gate bias. (c) Contact resistivity calculated by $\rho_c = R_c \times A$ for both 0.2 and 2 μm contact length. (d) Transfer length calculated from (a) and (c).

Now understanding how $L_T$ changes with gate voltage, we can now discuss the switching mechanism in MoS$_2$ transistors. From the discussion above we already understand why $L_T$ changes with gate bias. As we have mentioned above, $L_T = \sqrt{\rho_c / R_{sh}}$ is determined by both $\rho_c$ and $R_{sh}$. With increasing the gate bias, all three parameters, $L_T$, $\rho_c$ and $R_{sh}$, are decreasing. This indicates $\rho_c$ drops faster than $R_{sh}$ when increasing gate bias. From Figure 3(a) and 3(c) we can see, for the 2 μm contact transistor, a 100 V change in gate voltage results in a factor of ~30 decrease in sheet resistance, while the contact resistance decreases almost a factor of ~200. In other words, it is the contact rather than the channel that is more sensitive to the change in gate voltage. This means that the on/off switching in MoS$_2$ transistors are not primarily achieved by accumulating/depleting the carrier density in the channel, but by tuning the Schottky barrier width or the effective Schottky barrier height at source/drain junctions.[18,19] This is the fundamental difference between MoS$_2$ transistors and Si MOSFETs. At negative gate bias, the conduction band moves upwards, resulting in an enlarged effective Schottky barrier height for electrons, impeding carrier injection from the contact metal to MoS$_2$, which corresponds to the off-state of the device. On the other hand, at positive gate bias, the conduction band moves down. In spite of the absolute height for this barrier remaining intact, the narrowed barrier width facilitates thermally-assisted tunneling or even direct tunneling, thus the device is switched to the on-state. Because this switching mechanism is completely different from that of Si MOSFETs, the extraction of device parameters using the classical methods may not be appropriate. A typical example is the estimation of field-effect mobility in MoS$_2$. In Si MOSFETs, when the transistor is turned on, the surface potential and carrier density has no significant change. In the linear region, by using long channel approximation, the I-V characteristics can be written as: $I_{ds} = \mu_{eff} \frac{W}{L} C_{ox}(V_{gs} - V_{th})V_{ds}$, where $\mu_{eff}$ is the effective mobility, $W$ and $L$ are the channel width and length, $C_{ox}$ is gate oxide, $V_{gs}$, $V_{th}$ and $V_{ds}$ are the gate voltage, threshold voltage and drain voltage.[20] Once the transistor is turned on, the device can be modeled as a resistor with $R = 1/[\mu_{eff} \frac{W}{L} C_{ox}(V_{gs} - V_{th})]$. However, for MoS$_2$ transistors, even when the device is in the on-state, the transistor cannot be modeled as a resistor, since 1) we still have two Schottky barriers at the contact, and the effective height of the Schottky barrier changes with the $V_{gs}$, and 2) for the MoS$_2$ channel, the on-state in the transistor is mainly triggered by the reduced Schottky barriers, meanwhile the surface potential in MoS$_2$ channel may still be varied by gate bias. Therefore, we question the precision of field-effect mobility extraction simply from the transconductance peak in previous studies in light of the Schottky barriers discussed in this study. This explains why field effect mobility calculated from single- or few-layer MoS$_2$ transistors are lower than those values obtained from Hall mobility measurements.[21,22]

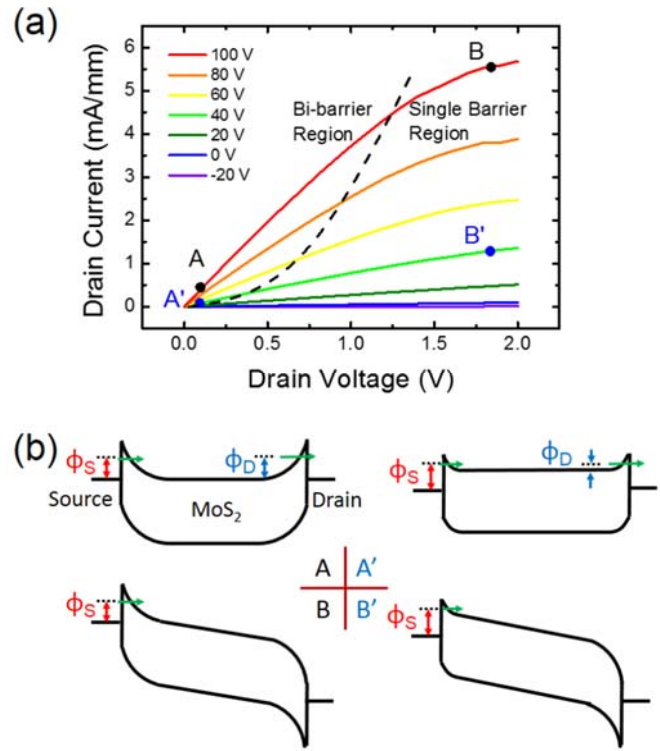

Figure 5: (a) Output characteristics of a 2 μm contact length transistor. $W$ and $L$ are 2 and 1.1 μm for this transistor. Back gate bias ranges from -20 V to 100 V with a 10 V step. The boundary between bi- and single-barrier regions is roughly indicated by the dashed line.(b) Band diagram according to the 4 biasing conditions denoted in (a).

In the previous paragraph we discussed how gate bias changes the effective barrier height and hence controls the switching in MoS$_2$ transistors. Now, we study further to understand how drain bias influences the Schottky barriers at drain and source. Here we limit our discussion to long channel devices only. In MoS$_2$ transistors, since we have two metal contacts that serve as source and drain, we have two Schottky barriers, the source barrier and the drain barrier. These two barriers are usually asymmetric. If we define the electron flow path from source to drain, the electrons would encounter the source barrier first, where they would undergo a thermal-assisted tunneling process from the metal Fermi level to the conduction band. On the other side of the transistor, a tunneling process may take place again depending on the drain bias of the device, and the electrons would go from the conduction band to the metal drain. In accordance with the long channel approximation, we use output curves measured from a 2 μm contact length transistor to illustrate the band diagrams at 4 different on-state bias conditions, as depicted in Figure 5. We first take a look at the high gate bias situation (point A and B), where gate is biased at 100 V. Both conduction and valence bands are pulled down, facilitating thermal-assisted tunneling from source metal to conduction band in MoS$_2$. At point A, when drain voltage is low, both source and drain barriers impede the electron transport. Note that the gate bias has an opposite impact on these two barriers. With higher gate bias, the effective barrier height for source barrier, $\Phi_S$, is reduced due to a sharper triangular barrier, meanwhile the barrier height for electrons injected from the semiconductor to the metal drain, $\Phi_D$, is enhanced due to the lowering of the conduction band, as compared to the low gate bias condition. However, with further increase of the drain bias, $\Phi_S$ remains constant, as it is fixed by the gate bias, however,

$\Phi_D$ keeps reducing and finally diminishes, leaving only one barrier and the device's diffusive channel, as shown in point B in Figure 5(b). As we have discussed previously, the modulation of the barrier heights is the dominate mechanism which changes the device conductance. At point A, since the increase in drain bias is lowering $\Phi_D$, the current has a sharper increase. After $\Phi_D$ is reduced to zero, the drain bias only acts on the device's diffusive channel so that current increase is not as fast at point B as it is at point A. This looks like current saturation in Si MOSFETs, but the difference is that the current saturation in $MoS_2$ transistors is caused by the changes in the barrier heights rather than pinch-off of the channel. For the lower gate bias situation (point A' and B'), where the gate bias is 40 V, the conduction band is not as low as the previous case, making $\Phi_D$ much lower. With increasing drain bias, $\Phi_D$ will be quickly reduced to zero. That is to say, in the lower gate bias case, the drain barrier is expected to have less of an impact on the output curves, making the $I_d$-$V_{ds}$ relationship look more linear. The barrier controlled output characteristics in $MoS_2$ transistors are completely reversed as compared to Si MOSFETs output curves. In n-type Si MOSFETs, the output curves at lower $V_{gs}$ saturate easier, where $V_{dsat}=V_{gs}-V_{th}$ is smaller. In the top curves, where $V_{gs}$ is higher, the output curves are usually straighter in the same drain bias range, as a higher $V_{ds}$ is needed for current saturation. However, in $MoS_2$ transistors, output curves at lower $V_{gs}$ look more linear while they are more curving at high $V_{gs}$. In summary, the device performance in long channel $MoS_2$ transistors is simply controlled by two Schottky barriers. The gate voltage controls both barriers to switch the device between on- and off-states, while the drain voltage has a larger impact on the drain barrier.

Through its action on the drain barrier, the increase in $V_{ds}$ shifts the device from the "bi-barrier region" to the "single-barrier region" in the on-state, similar to the linear and saturation regions of Si MOSFETs.

Finally, we would like to briefly discuss the MoS$_2$ transistors at short channel regions, to understand how these barriers influence the device performance. The existence of barriers is not desirable, as it introduces a large contact resistance which limits the on-current in the transistor. However, when channel length is aggressively reduced and it is comparable to the barrier width, the drain bias would influence $\Phi_S$ as well. In the on-state, an increased drain bias would reduce the source barrier width, hence reducing the source contact resistance. It is similar to drain-induced barrier lowering (DIBL) in short channel Si MOSFETs, where the drain bias lowers the barrier in the channel at off-state, making the device difficult to turn off. Hence, in the Si MOSFET case, DIBL is undesirable as it degrades the off-state of the transistors. However, in MoS$_2$ transistors, the drain-induced barrier narrowing (DIBN) would be a "favorable" short channel effect as it reduces the Schottky barrier width and enhances the on-state current. This means that one of the bottlenecks for MoS$_2$ transistors, the large contact resistance, may act only in long channel regions and diminish at the short channel length. Therefore, short channel MoS$_2$ transistors would be potentially a competitive technology once the channel length is aggressively scaled down to the level of barrier width. The source barrier can be increased in the off-state to impede current flow; however the barrier width can be narrowed at larger drain bias to increase on-state current, due to the DIBN effect. In other words, the contact resistance at shorter channel length would be significantly reduced, making contact dimensions scalable as well.

**CONCLUSIONS**

In summary, we study the $MoS_2$ transistor behavior with various contact lengths. We find the device performance is strongly related to the contact dimensions. Sheet resistance, contact resistivity and transfer lengths at various gate voltages are extracted. We reveal that the switching and output behaviors of $MoS_2$ transistors are modulated by two Schottky barriers. The $MoS_2$ transistor would potentially be a very promising device at short channel regions after optimizing the device design.

**METHODS**

Single layer $MoS_2$ crystals were achieved from chemical vapor deposition process on heavily doped silicon wafer (0.01-0.02 Ω·cm) with 285 nm $SiO_2$ on top. A 1 min dry etching was used to pattern the $MoS_2$ crystals with $BCl_3$/Ar plasma. The flow rate was 15 and 60 sccm for $BCl_3$ and Ar. The RF source power and RF bias were 100 W and 50 W, respectively. Metal contacts were defined by electron beam lithography, followed by the electron beam evaporation of Ti/Au for 20/60 nm with the deposition rate of ~1 Å/s. Electrical characterizations were carried out with Keithley 4200 Semiconductor Characterization System at room temperature.

**ACKNOWLEDGEMETNS**

The authors would like to thank Prof. Kosuke Nagashio for valuable discussions. This material is based upon work partly supported by NSF under Grant CMMI-1120577 and SRC under Tasks 2362 and 2396. S. N., P. M. A. and J. L. acknowledge the support of Welch Foundation grant C-1716, the NSF grant ECCS-1327093 and the U.S. Army Research Office MURI grant W911NF-11-1-0362.

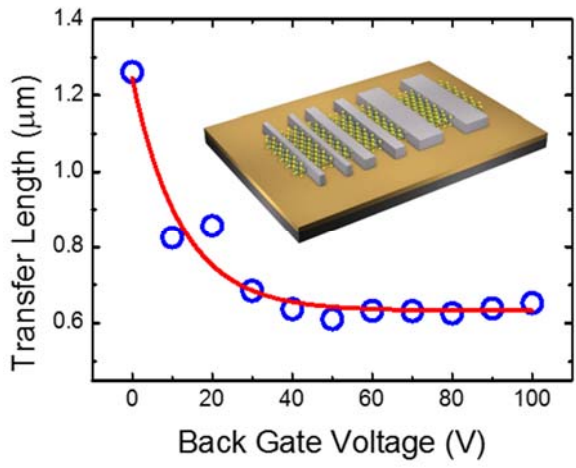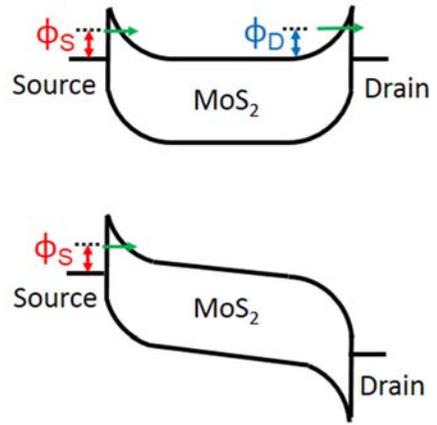

**TOC Graphics**